\documentclass{article}

     \PassOptionsToPackage{numbers, compress}{natbib}
\usepackage{ifthen}
\newboolean{arxiv}
\setboolean{arxiv}{false}
\ifthenelse{\boolean{arxiv}}{
  \usepackage[preprint]{neurips_2024}
}{
  \usepackage[final]{neurips_2024}
}

\usepackage[utf8]{inputenc} % allow utf-8 input
\usepackage[T1]{fontenc}    % use 8-bit T1 fonts
\usepackage[hidelinks]{hyperref}       % hyperlinks
\usepackage{url}            % simple URL typesetting
\usepackage{booktabs}       % professional-quality tables
\usepackage{amsfonts}       % blackboard math symbols
\usepackage{nicefrac}       % compact symbols for 1/2, etc.
\usepackage{microtype}      % microtypography
\usepackage{xcolor}         % colors
\usepackage{amsmath}    % <----- don't forget
\usepackage[capitalise, noabbrev]{cleveref}
\usepackage{enumitem}
\usepackage{graphicx}
\usepackage{verbatim}
\usepackage[export]{adjustbox}
\newcommand{\st}{\sffamily}
\graphicspath{{./Figures/}}

\usepackage{algorithm,algpseudocode}
\algnewcommand{\Inputs}[1]{%
  \State \textbf{Inputs:}
  \Statex \hspace*{\algorithmicindent}\parbox[t]{.8\linewidth}{\raggedright #1}
}
\algnewcommand{\Initialize}[1]{%
  \State \textbf{Initialization:}
  \Statex \hspace*{\algorithmicindent}\parbox[t]{.8\linewidth}{\raggedright #1}
}

\renewcommand\citet[1]{\citeauthor{#1} (\citeyear{#1})}

\newcommand{\tightlist}[1][0pt]{%
  \setlength{\itemsep}{#1}%
  \setlength{\parskip}{0pt}
}

\title{Autoregressive Image Diffusion: Generation of Image Sequence and Application in MRI}

\ifthenelse{\boolean{arxiv}}{
  \usepackage{authblk}

  \setlength{\affilsep}{4pt}
  \author[1]{Guanxiong Luo}
  \author[3]{Shoujin Huang}
  \author[1,2]{Martin Uecker}  
  \affil[1]{Institute for Diagnostic and Interventional Radiology, University Medical Center G\"ottingen, Germany}
  \affil[2]{Institute of Biomedical Imaging, Graz University of Technology, Graz, Austria}
  \affil[3]{Shenzhen Technology University, Shenzhen, China}
}{
  \author{%
  Guanxiong Luo\\
  University Medical Center G\"ottingen\\
  guanxiong.luo@med.uni-goettingen.de\\
  \And
  Shoujin Huang\\
  Shenzhen Technology University\\
  \And
  Martin Uecker\\
  Graz University of Technology\\
  uecker@tugraz.at
  }
}

\begin{document}
\maketitle
\begin{abstract}
Magnetic resonance imaging (MRI) is a widely used non-invasive imaging modality. However, a persistent challenge  lies in balancing image
quality with imaging speed. This trade-off is primarily constrained by k-space measurements, which traverse specific trajectories in the spatial Fourier domain (k-space). These measurements are often undersampled to shorten acquisition times, resulting in image artifacts and compromised quality. Generative models learn image distributions and can be used to reconstruct high-quality images from undersampled k-space data. 
In this work, we present the autoregressive image diffusion (AID) model for image sequences and use it to sample the posterior for accelerated MRI reconstruction. The algorithm incorporates both undersampled k-space and pre-existing information. Models trained with fastMRI dataset are evaluated comprehensively.
The results show that the AID model can robustly generate sequentially coherent image sequences. In MRI applications, the AID can outperform the standard diffusion model and reduce hallucinations, due to the learned inter-image dependencies. The project code is available at \url{https://github.com/mrirecon/aid}.
\end{abstract}

\section{Introduction}

Magnetic resonance imaging (MRI) is a non-invasive imaging modality widely used in clinical practice to visualize soft tissue. Despite its utility, a persistent challenge in MRI is the trade-off between image quality and imaging speed. The trade-off is influenced by the k-space (spatial Fourier domain) measurements, which traverse spatial frequency data points along given sampling trajectories. To reduce acquisition time, the k-space measurements are often undersampled, resulting in image artifacts and reduced image quality.

In recent years, deep learning-based methods have emerged to improve image reconstruction in MRI. These methods are formulated as an inverse problem building upon compressed sensing techniques \cite{Lustig_Magn.Reson.Med._2007, Block_Magn.Reson.Med._2007} and benefit from the learned prior information instead of hand-crafted priors \cite{Yang_NIPS_2016, Hammernik_Magn.Reson.Med._2017, Mardani_IEEETrans.Med.Imag._2018}. Another successful approach involves learning an image prior parameterized by a generative neural network \cite{Tezcan_IEEETrans.Med.Imag._2018, Luo_Magn.Reson.Med._2020}, which is then used as the learned and decoupled regularization on the image. Generative priors offer flexibility in handling changes in the forward model and perform well in reconstructing high-quality images from undersampled data.

Diffusion models \cite{sohl2015deep, Song_NIPS_2019, ho2020denoising}, a class of generative models, have gained attention in recent years and are making an impact in many fields, including MRI reconstruction \cite{jalal2021robust,yang2023diffusion}. These models learn to reverse a diffusion process that transforms random noise into structured images, producing high-quality, detailed images. Various approaches, including denoising diffusion probabilistic models (DDPMs) \cite{ho2020denoising}, denoising score matching \cite{Song_NIPS_2019}, and continuous formulations based on stochastic differential equations (SDEs) \cite{Song_ICLR_2021}, have been proposed for deriving diffusion models.

Recent studies demonstrate the effectiveness of diffusion models in accelerated MRI and their flexibility in handling various sampling patterns \cite{jalal2021robust, gungor2023adaptive, chung2022score, luo2023bayesian, zach2023stable}. For example, training score-based generative models using Langevin dynamics yields competitive reconstruction results for both in-distribution and out-of-distribution data \cite{jalal2021robust}. Additionally, score-based diffusion models trained solely on magnitude images can reconstruct complex-valued data \cite{chung2022score}. Comprehensive approaches using data-driven Markov chains facilitate efficient MRI reconstruction across variable sampling schemes and enable the generation of uncertainty maps \cite{luo2023bayesian}.

Autoregressive models are statistical models that predict the current value of a variable based on its past values, capturing temporal dependencies and patterns within the data. They are widely used in various fields such as time series analysis, signal processing, and sequence modeling. In natural language processing, autoregressive models like generative pre-trained transform (GPT) \cite{vaswani2017attention, radford2018improving} predict each token in a sequence based on previously generated tokens, enabling the generation of coherent and contextually relevant text. Similarly, in image modeling, autoregressive models like PixelCNN \cite{van2016conditional} and ImageGPT \cite{chen2020generative} generate images by predicting each pixel value based on previously generated pixel values, often in a left-to-right, top-to-bottom order. Instead of directly modeling pixels, which can be computationally expensive for high-resolution images, the study \cite{esser2021taming} proposes to first compress the image into a smaller representation using vector quantized variational autoencoder (VQVAE). This VQVAE learns a codebook of visually meaningful image components. Then, a transformer is applied to model the autoregressive relationship between these components, effectively capturing the global structure of the image. By predicting each image component based on previous ones, the model generates high-resolution images in a sequential manner, maintaining consistency and coherence across the entire image.

The clinical practice of MRI often involves acquiring volumetric image sequences to monitor disease progression and treatment response; modeling and generating these image sequences is challenging.
Autoregressive models can be employed to model the joint distribution of image sequences and extract the dependencies between images.
The diffusion process is effective in modelling images by treating each image independently. 
Therefore, we aim to combine these two models and propose autoregressive image diffusion (AID) model to generate sequences of images. 

The contributions of this work are the following aspects. We present how to derive the autoregressive image diffusion training loss starting from a common diffusion loss and how to optimize loss in parallel for efficient training. We present the algorithm to sample the posterior for accelerated MRI reconstruction when using AID to facilitate the incorporation of pre-existing information. We performed experiments to evaluate its ability in generating images when different the amount of initial information is given and to validate its effectiveness in MRI reconstruction.
The results show that the AID model can stably generate highly coherent image sequences even without any pre-existing information.  When used as a generative prior in MRI reconstruction, the AID outperforms the standard diffusion model and reduces the hallucinations in the reconstructed images, benefiting from the learned prior knowledge about the relationship between images and pre-existing information.

\begin{figure}
  \centering
  \includegraphics[width=0.7\textwidth]{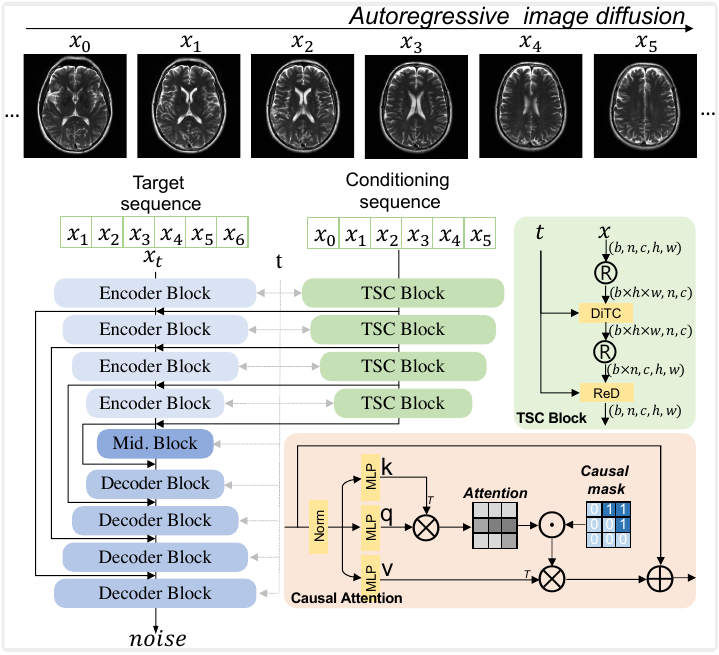}
  \caption{The interaction between the images in conditioning sequence occurs in the DiTBlock, which has a causal attention module to ensure $x_n$ is conditioned on previous images $x_{<n}$. During training, the net predicts the noise for each noisy image that is sampled from the target sequence given the conditioning sequence in parallel. %FIXME, this second sentence is unclear
During generation, the net iteratively refines the noisy input to produce a clean image, which is then appended to the conditioning sequence.}
  \label{fig:arch}
\end{figure}

\section{Methods}
\subsection{Autoregressive image diffusion}
Given a dataset \( X \) consisting of multiple sequences of images, each sequence represented as \( \mathbf{x} = \{x_1, x_2, \ldots, x_N\} \), our goal is to model the joint distribution of these images. This joint distribution is  autoregressively factorized into the product of conditional probabilities:
\begin{equation}
p(\mathbf{x}) = q(x_1|x_0) \prod_{t=2}^N q(x_n|x_{<n}),
\end{equation}
where \( x_{<n} = \{x_1, x_2, \ldots, x_{n-1}\} \) and the image $x_0$ is known. The model parameterized by $\theta$ is trained by minimizing the negative log-likelihood of the data:
\begin{equation}
\mathcal{L}_{AID} = \mathbb{E}_X\left[-\log p_\theta(\mathbf{x})\right] = \mathbb{E}_X\left[-\log p_\theta(x_1|x_0) - \sum_{t=2}^N \log p_\theta(x_n|x_{<n})\right].
\label{eq:ob_aid}
\end{equation}

\citet{sohl2015deep} and \citet{ho2020denoising} introduced the denoising diffusion probabilistic model (DDPM). This model gradually introduces fixed Gaussian noise to an observed data point \( x^0 \) using known scales \( \beta_t \), generating a series of progressively noisier values \( x^1, x^2, \ldots, x^T \). The final noisy output \( x^T \) follows a Gaussian distribution with zero and identity covariance matrix \( I \), containing no information about the original data point. The series of positive noise scales \( \beta_1, \ldots, \beta_T \) must be increasing, ensuring that the first noisy output \( x^1 \) closely resembles the original data \( x^0 \), while the final value \( x^T \) represents pure noise. 
We apply this process to the conditional probability $q(x_n|x_{<n})$ in \cref{eq:ob_aid} by adding the noise to the image independent of the position in the sequence, i.e., $x_n^t$ and $x_{<n}^0$ are conditionally independent given $x_n^{t-1}$. Then the transition from $x_n^{t-1}$ to $x_n^t$ is defined as:
\begin{equation}
  q(x_n^t|x_n^{t-1}, x_{<n}^0) = q(x_n^t|x_n^{t-1}) = \mathcal{N}(x_n^t; \sqrt{1-\beta_t}x_n^{t-1}, \beta_t\mathbf{I})  
\end{equation}
Here, \( x_n^t \) represents the image \( x_n \) at time \( t \), \( x_n^{t-1} \) is the image at the previous time step, and \( x_{<n}^0 \) denotes all images preceding \( x_n \) at the initial time step. The parameter \( \beta_t \) controls the drift and diffusion of this process.
The objective is to learn to reverse this process. The reverse process is defined as:
\begin{equation}
  p_\theta(x_n^{t-1}|x_n^t, x_{<n}^0) = \mathcal{N}(x_n^{t-1}; \mu_\theta(x_n^t, x_{<n}^0, t), \Sigma_\theta(x_n^t, x_{<n}^0, t)),  
  \label{eq:reverse_a}
\end{equation}
where \( \mu_\theta \) and \( \Sigma_\theta \) are parameterized by a neural network \( \theta \), taking \( x_n^t \), \( x_{<n}^0 \), and \( t \) as inputs. Using the variational lower bound, the reverse process can be learned by minimizing the negative log-likelihood of the data:
\begin{equation}
  \mathbb{E}[-\log p_\theta(x_n|x_{<n}^0)] \leq \mathbb{E}\left[-\log p(x_n^T)-\sum_{t\geq1}\log \frac{p_\theta(x_n^{t-1}|x_n^t, x_{<n}^0)}{q(x_n^t|x_n^{t-1}, x_{<n}^0)}\right] := L_{D_n},
  \label{eq:dn}
\end{equation}

Given the initial image \( x_n^0 \) and that $x_n^t$ and $x_{<n}^0$ are conditionally independent given $x_n^0$, $x_n^t$  at an arbitrary time step \( t \) is sampled from a Gaussian distribution:
\begin{equation}
  q(x_n^t| x_n^0, x_{<n}^0) = \mathcal{N}(x_n^t; \sqrt{\bar{\alpha}_t}x_n^0, (1-\bar{\alpha}_t)\mathbf{I}),
\end{equation}
using \( {\alpha}_t=1-\beta_t \) and \(\bar{\alpha}_t=\prod_{s=1}^t\alpha_s\).
The posterior distribution \( x_n^{t-1} \) given \( x_n^0 \) and \( x_n^t \) is then calculated as:
\begin{equation}
  q(x_n^{t-1}|x_n^t, x_n^0, x_{<n}^0) = \mathcal{N}(x_n^{t-1}; \tilde{\mu}_t(x_n^t, x_n^0), \tilde{\beta}_t\mathbf{I}),
  \label{eq:po}
\end{equation}
where $\tilde{\mu}_{t}({x}^{t}_n,{x}_{n}^0):={\frac{\sqrt{\alpha_{t-1}}\beta_{t}}{1-\bar{\alpha}_{t}}}{x}_{n}^0+{\frac{\sqrt{\alpha_{t}}(1-\bar{\alpha}_{t-1})}{1-\bar{\alpha}_{t}}}{x}_{n}^t$ and $\tilde{\beta}_{t}:={\frac{1-\tilde{\alpha}_{t-1}}{1-\tilde{\alpha}_{t}}}\beta_{t}$.

The training objective \cref{eq:dn} is further written as minimizing the Kullback-Leibler (KL) divergence between the forward and reverse processes in \cref{eq:reverse_a} and \cref{eq:po}, as proposed by \citet{sohl2015deep}. (See \cref{ap:loss} for details.) 

In practice, the approach proposed by \citet{ho2020denoising} involves reparameterizing \( \mu_{\theta} \) and predicting the noise \( \epsilon \) for \( x_n^t \). The expression for \( x_n^t \) is given by \( x_n^t(x_n^0, \epsilon) = \sqrt{\bar{\alpha}_t}x_n^0 + \sqrt{1 - \bar{\alpha}_t}\epsilon \), with \(\Sigma_\theta(x_n^t, x_{<n}^0, t) = \beta_t\) fixed. We realized this with a neural network \( \epsilon_{\theta}(x_n^t, t, x_{<n}^0) \) shown in \cref{fig:arch},  which predicts the noise for $x_n^t$ at each time step given $x_{<n}^0$. In the end, the objective function in \cref{eq:ob_aid} for training autoregressive image diffusion is written as 
\begin{equation}
  \mathcal{L}_{AID} \geq \sum_{n=1}^{N} L_{D_n} = \sum_{n=1}^{N}\mathbb{E}_{t, \epsilon | x_n^0, x_{<n}^0} \left[ \left\| \epsilon_\theta (\sqrt{\bar{\alpha}_t} x_n^0 + \sqrt{1-\bar{\alpha}_t} \epsilon, x_{<n}^0, t) - \epsilon \right\|_2^2 \right], 
  \label{eq:aid}
\end{equation}
where the expectation is taken over the noise \( \epsilon \sim \mathcal{N}(0, I) \) and the time step \( t \sim \mathcal{U}({1, ..., T}) \). To generate an image sequence, we begin with the noise \( x_1^T \) and update it iteratively using \cref{eq:reverse_a} with the given \( x_0^0 \), following the sequence \( (x_1^T \rightarrow x^{T-1} \rightarrow \ldots \rightarrow x_1^0) \). This process yields a clean sample \( x_1^0 \). Subsequently, we can sample \( x_2^0 \) in the same manner using the generated images \( x_{<2}^0 \), and continue this process iteratively to generate the entire sequence of images.

\subsection{Architecture}
To optimize the objective function in \cref{eq:aid} efficiently, ordered images are
loaded as sequences of a certain length \( N+1 \) during the training phase.
We take the first \( N \) images \(\mathbf{x}_{con}=\{x_0,x_1,...,x_{N-1}\}\) 
as the conditioning sequence and the last \( N \) images \(\mathbf{x}_{target}=\{x_1, ...., x_N\}\) as the target sequence, as shown in \cref{fig:arch}.
We adopt an architecture built on an Unet \cite{ronneberger2015u} with capabilities of temporal-spatial conditioning (TSC), designed to process the conditioning sequence and predict the noise for the target sequence. The term "temporal" refers to conditioning in previous frames along the \( N \) dimensions, while the "spatial" refers to the conditioning in the previous frame among the \( H \times W \) dimensions. Additionally, the TSC block is conditioned on the time steps \( t \) of the diffusion process.

The only interaction between images in the conditioning sequence occurs during the attention operation. To maintain proper conditioning with autoregressive property, we implemented a standard upper triangular mask on the $n\times n$ matrix of attention logits. This causal attention module is used in DiTBlock \cite{vaswani2017attention, peebles2023scalable}. The modified DiTBlock is followed by a ResNet block \cite{he2016deep}, which is a standard building block in the Unet architecture. The features output by the TSC block are then passed to the corresponding encoder block in the Unet, which process the target sequence. The change in tensor dimensions inside TSC Block is handled by the einops library\footnote{https://github.com/arogozhnikov/einops} and illustrated in \cref{fig:arch}.

During training, the net predicts the noise in parallel for each noisy image that is sampled from the target sequence, given the conditioning sequence. During generation of sequence, the net iteratively refines the noisy input to produce a clean image, which is then appended to the conditioning sequence.

\subsection{Application in MRI inverse problem}
Image reconstruction is formulated as a Bayesian problem where the posterior of image $p(x|y)$ is expressed as
\begin{align}
	p({x}|{y}) &= \frac{p({y}|{x})\cdot p({x})}{p({y})} ~.
	\label{eq:bayes}
\end{align}
Here, $y$ represents the measured k-space data, $x$ denotes the image, and $p(x)$ is a generative prior.
The minimum mean square error (MMSE) estimator for the posterior minimizes the mean square error, given by:
\begin{equation}
	{{x}}_\mathrm{MMSE} = \arg \min_{\tilde{{x}}} \int \|\tilde{{x}} - {x}\|^2 p({x}|{y})d{x} = \mathbb{E}[x|y]~.
  \label{eq:mmse}
\end{equation}

\subsection{Likelihood function for k-space}
The image \( x \in  \mathbb{C}^{n \times n}  \)  is represented as a complex matrix , where \( n \times n \) is the image size, and \( y \in \mathbb{C}^{m \times m_C} \) is a vector of complex-valued k-space samples from \( m_C \) receive coils. Assuming circularly-symmetric normal noise \( \eta \) with zero mean and covariance matrix \( \sigma^2_{\eta} \mathbf{I} \), the likelihood \( p(y|x) \) of observing \( y \) given \( x \) is formulated as a complex normal distribution:
\begin{align}
	p({y}|{x}) & = \mathcal{CN}({y}; \mathcal{A}{x}, \sigma^2_{\eta} \mathbf{I}) \nonumber\\
	& = (\sigma_{\eta}^2\pi)^{-N_p} e^{\text{-}\|\sigma_{\eta}^{-1} \cdot ({y} - \mathcal{A}{x})\|_2^2}~,
	\label{eq:forw} 
\end{align}
where \( \mathbf{I} \) is the identity matrix, \( \sigma_{\eta} \) is the standard deviation of the noise, \( \mathcal{A}x \) represents the mean, and \( N_p \) is the length of the k-space data vector. The operator \( \mathcal{A}:\mathbb{C}^{n \times n}\rightarrow\mathbb{C}^{m \times m_C} \) maps the image \( x \) to k-space and is composed of the coil sensitivity maps \( \mathcal{S} \), the two-dimensional Fourier transform \( \mathcal{F} \), and the k-space sampling mask \( \mathcal{P} \), defined as \( \mathcal{A} = \mathcal{PFS} \). For more details and visual understanding on the forward operator, please refer to \cref{ap:lik}.

\subsection{Sampling the posterior}
Given a sequence of k-space \(\mathbf{y}=\{y_1,\ldots, y_N\}\), each posterior in \( \{p_\theta(x_n|y_n, x_{<n}^0 )| 1<n<N\} \) is expressed as
\begin{align}
p_\theta(x_n|y_n, x_{<n}^0) & = \frac{p(y_n|x_n, x_{<n}^0)p_\theta(x_n|x_{<n}^0)}{p(y_n|x_{<n}^0)} = \frac{p(y_n|x_n)p_\theta(x_n|x_{<n}^0)}{p(y_n)} \nonumber\\
 &\propto p(y_n|x_n)p_\theta(x_n|x_{<n}^0)~,
 \label{eq:pos1}
\end{align}
when the acquisition of $y_n$ is independent of the image $x_{<n}^0$, $y_n$ and $x_{<n}^0$ are conditionally independent given $x_n$.
Following the Reference \cite{sohl2015deep}, we have
\begin{equation}
  p_\theta(x_n^{t-1}|x_n^t, y_n, x_{<n}^0) \propto p(y_n|x_n^t)p_\theta(x_n^{t-1}|x^t_{<n},x_{<n}^0)~.
  \label{eq:pos}
  \end{equation}
The details for \cref{eq:pos} is in \cref{ap:post}. To sample the above posterior, the learned reverse process in \cref{eq:reverse_a} is used, and 
the algorithm is constructed with two gradient updates using the log of the prior and k-space likelihood: the DDIM (Denoising Diffusion Implicit Model) reverse step proposed by \citet{song2020denoising}, and a data fidelity step derived from the likelihood function \cref{eq:forw}, which are described as follows:
\begin{align}
  \tilde{x}_n^{t-1} &\leftarrow \sqrt{\alpha_{t-1}} \left( \frac{x_n^t - \sqrt{1 - \alpha_t} \epsilon_{\theta}(x_n^t,x_{<n}^0, t)}{\sqrt{\alpha_t}} \right) + \sqrt{1 - \alpha_{t-1}} \epsilon_{\theta}(x_n^t,x_{<n}^0, t) \label{eq:up1} \\
  x_n^{t-1}  &\leftarrow \tilde{x}_n^{t-1} + \lambda\cdot\nabla_{{x}_n^{t-1}} \log p(y_n|\tilde{x}_n^{t-1}) ~.
  \label{eq:up2}
\end{align} 
where $\lambda$ is the step size, and $\nabla_{{x}_n^{t-1}} \log p(y_n|x_n^{t-1})$ is the gradient of the log-likelihood of \cref{eq:forw}. Then, the reconstruction of a sequence images from the undersampled k-space data is achieved by sequentially sampling the posterior in \( \{p(x_n|y_n, x_{<n}^0 )| 1<n<N\} \) using autoregressive diffusion model as prior. The algorithm is summarized in Algorithm \ref{alg:seq}.
\begin{algorithm}
  \caption{Sample the posterior in \( \{p(x_n|y_n, x_{<n}^0 )| 1<n<N\} \) using autoregressive diffusion model as prior.}
  \label{alg:seq}
  \begin{algorithmic}[1]
  \State Initial image sequence: $x_{<n}^0 = x_0$;
  Time steps: $T$;
  Step size: $\lambda$;
  Iterations for data fidelity step: $K$;
  Number of samples: $S$;
  \For {$y_n$ in $\mathbf{y}=\{y_1, y_2, ..., y_N\}$ }
  \State Initialize $x_n^T$ with Gaussian noise.
  \State Construct the forward operator $\mathcal{A}$ with sampling pattern $\mathcal{P}$ and coil sensitivities $\mathcal{S}$.
  \For {$t$ in $\{T-1,\ldots,0\}$ }
  \State Run the DDIM reverse step in \cref{eq:up1} to get $x_n^{t-1}$ given $x_n^t$ and $x_{<n}^0$.
  \State Run the data fidelity step in \cref{eq:up2} to update $x_n^{t-1} $ for $K$ step.
  \State Add Gaussian noise scaled by \(\sqrt{1 - \alpha_{t-1}}\) to $x_n^{t-1}$.
  \EndFor
  \State Update $x_{<n}^0 \leftarrow \{x_n^0, \ldots, x_0^0\}$.
  \EndFor
  \end{algorithmic}
  
  \end{algorithm}
\section{Experiments and Results}

\subsection{Model training}
Two autoregressive diffusion models were trained on separate datasets: one in image space and the other in latent space. The image space model was trained on brain images that are from the fastMRI training dataset, which includes T1-weighted (some with post-contrast), T2-weighted, and FLAIR images \cite{knoll2020fastmri}. These complex images were reconstructed from fully sampled multi-channel k-space volumes, with coil sensitivity maps computed using the BART toolbox \cite{blumenthal_2023_10277939}. The images were then normalized to a maximum magnitude of 1, and the real and imaginary parts were treated as separate channels when input into the neural network. The number of images in each volume ranged from 12 to 16. Images were loaded without reordering and resized to 320$\times$320 pixels if they were not already that size.

The latent space model is trained with the cardiac dataset that contains cine images reconstructed by the SSA-FARY method \cite{rosenzweig2020cardiac}. Firstly, a VQVAE was trained on the cine images that were preprocessed similarly to images in fastMRI dataset. The cine images have a size of  256$\times$256 pixels. Then, it generates latent space for the training AID. (See the details for configuration of VQVAE in \cref{ap:vqvae}). All the training was performed on 4 NVIDIA A100 GPUs with 80GB memory. The models were trained using the Adam optimizer with a learning rate of \(10^{-4}\) and a batch size of 1 for image space model and 4 for latent space model. Two models were trained for 440,000 iterations. It took around 2 hours to train brain model for 10k steps and 1.2 hours for cardiac model. The length of conditioning sequence $N$ for brain and cardiac models are 10 and 42.
The network as illustrated in \cref{fig:arch} was implemented based on OpenAI's guided diffusion codebase\footnote{https://github.com/openai/guided-diffusion}.
We also trained a standard diffusion model, Guide, on the brain dataset for comparison. The Guide model was trained using the same hyperparameters as the AID model, except the batch size is 10. The Guide model uses the same Unet blocks as AID.

\subsection{Generating sequence of images}
To test different aspects of the autoregressive diffusion models, we generate the sequence of images using the following two approaches.

\textbf{Retrospective sampling}:
This method generates a new sequence of images $\{\tilde{x}_1, \ldots, \tilde{x}_{N}\}$ based on the given sequence $\{x_0, \ldots, x_{N-1}\}$. $\tilde{x}_n$ is sampled from \cref{eq:reverse_a} given $\{x_0, \ldots, x_{n-1}\}$.
\begin{figure}[htbp]
\resizebox{\textwidth}{!}{
\begin{tabular}{c@{\hskip 0pt} c}
    {\st (a)}&\begin{tabular}{c}
      \includegraphics[width=\textwidth]{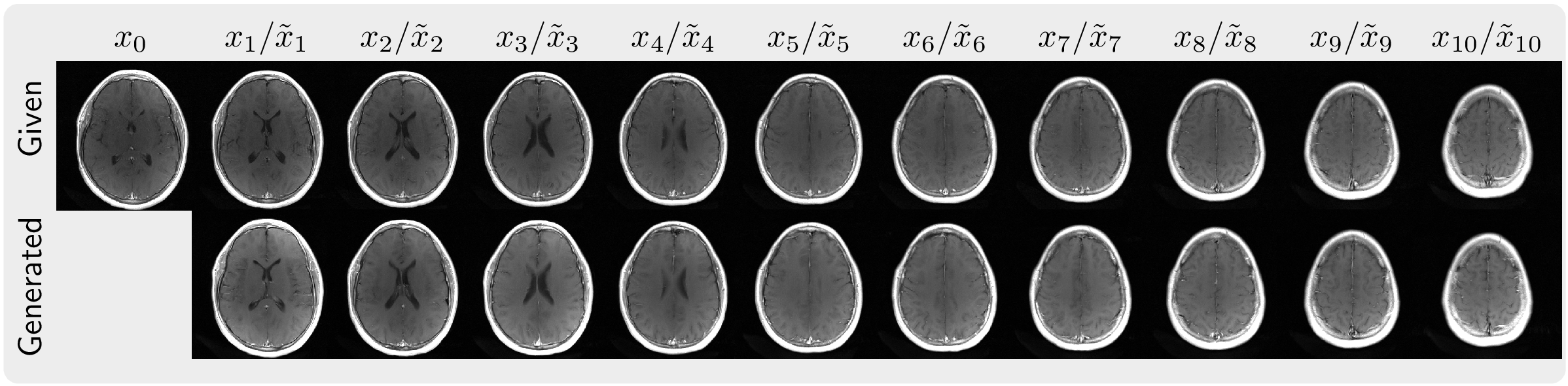}
    \end{tabular}\\
    {\st (b)}&\begin{tabular}{c}
      \includegraphics[width=\textwidth]{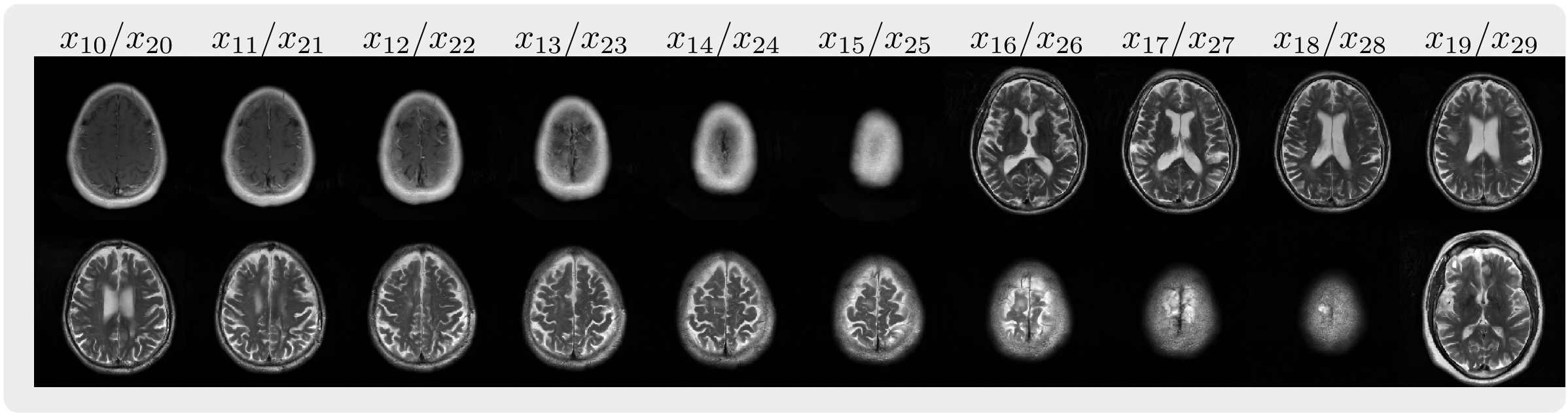}
    \end{tabular}
  \end{tabular}}
  
 \caption{(a): A sequence of images from dataset is shown in the first row and is used as conditioning to generate retrospective samples that are shown in the second row. (b): With the given sequence in (a) as a warm start, prospective samples extending it are shown.}
  \label{fig:retro}
\end{figure}
\begin{figure}[htbp]
  \centering
  \includegraphics[width=0.95\textwidth]{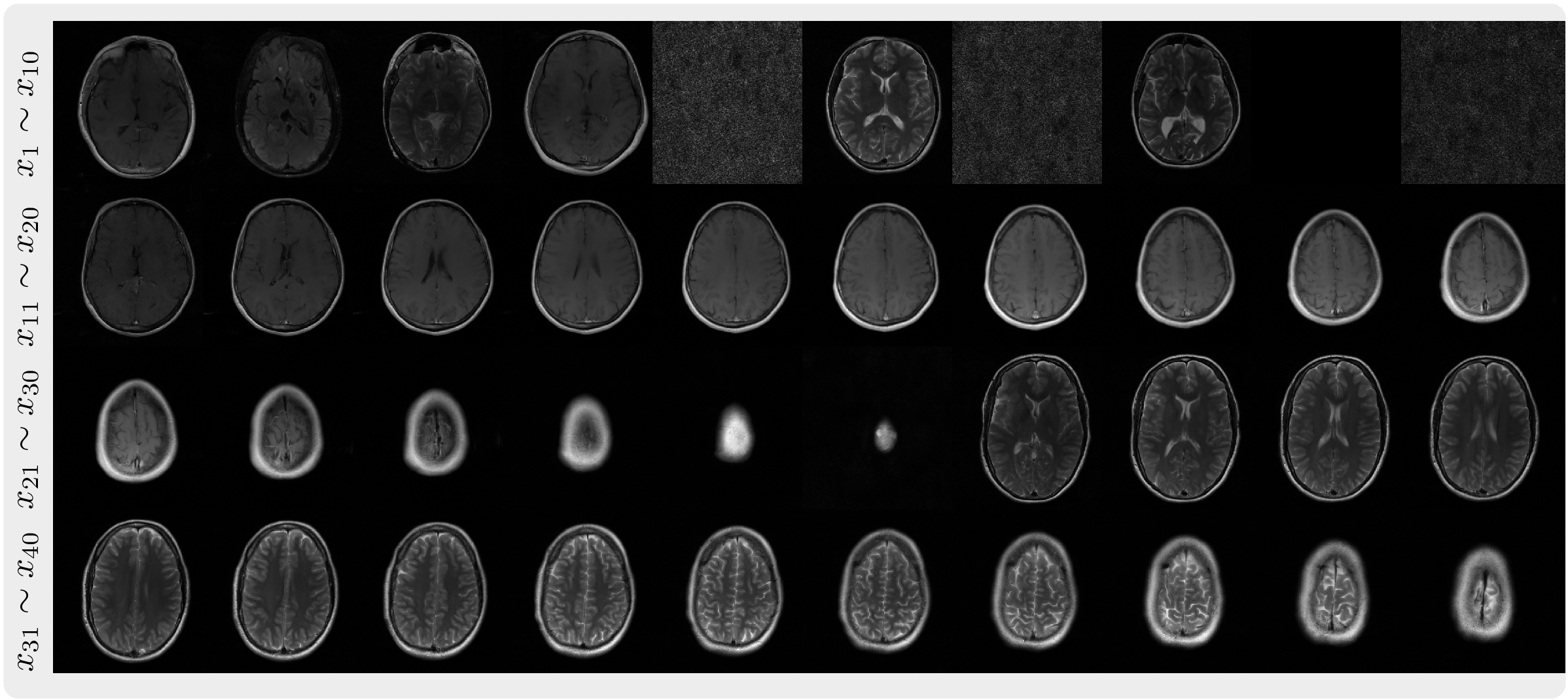}
  \caption{Prospective samples with cold start. The initial images generated in the cold start are not sequentially coherent, but as the sampling process continues, the model progressively generates more sequentially coherent and realistic images.}
  \label{fig:cold}
\end{figure}

\textbf{Prospective sampling}:
A fixed-length sliding window is initialized with the given sequence $x_{<N}=\{x_0, \ldots, x_{N-1}\}$. \(x_N\) is generated from \cref{eq:reverse_a} with the current window as conditioning. Subsequently, the window is updated by appending the newly generated $x_N$ and removing the earliest image $x_0$. This autoregressive sampling process is repeated until the stop condition is met. We refer to this process as a warm start. In a cold start, the window is initialized with zeros, and each element $x_n$ in it is updated with newly generated images from the beginning to the end, after which the generation is warmed up.

In the retrospective sampling, the model generates a sequence of images that are sequentially coherent and visually similar to the conditioning sequence, as shown in \cref{fig:retro} (a). The prospective sampling generates a sequence of images that extends the initial images in the sliding window and constitutes multiple volumes, as shown in \cref{fig:retro} (b). 
As for a cold start, \cref{fig:cold} demonstrates the model's ability to generate a sequence of images using black background as initial status. 
This shows the model's generative capabilities from a minimal initial condition, thereby proving its robustness and flexibility.
Due to the limit of space, the samples with similar quality from the model trained on the cardiac dataset are shown in \cref{ap:cardiac}. We also implemented a boosted sampling technique which use previous slice with added noise as the initial image for the current slice. This requires less iterations to generate the sequence of images. Further details can be found in our codebase.

\subsection{MRI reconstruction}

The MMSE estimator in \cref{eq:mmse} cannot be computed in a closed form, and numerical approximations are typically required. Once the samples from the posterior is obtained with \cref{alg:seq}, a consistent estimate of ${x_n}_\mathrm{MMSE}$ can be computed by  averaging those samples, i.e. the empirical mean of samples converges in probability to ${x_n}_\mathrm{MMSE}$ due to weak law of large numbers.
The variance of those samples provides a solution to the error assessment in the reconstruction assuming the trained model is trusted. To highlight the regions with large uncertainty, we compute the pseudo-confidence intervals based on the assumption that each pixel's intensity is normally distributed.
This involves determining the standard error from the variance, then multiplying it by the t-score corresponding to a 95\% confidence level.

\begin{figure}[htbp]
  \centering
  \includegraphics[width=0.8\textwidth]{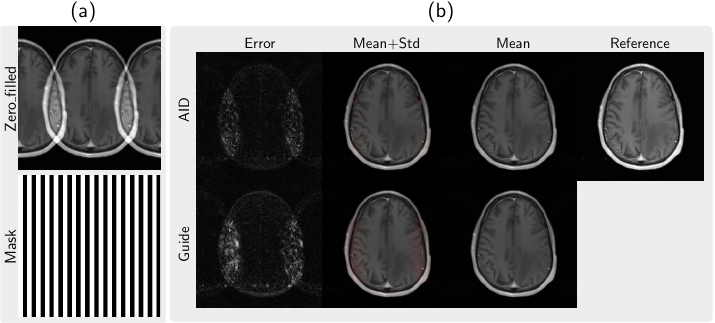}
  \caption{(a): The folded single-coil image caused by two-times undersampling mask. (b): The comparison of unfolding ability by the autoregressive and the standard diffusion model, i.e., AID (top) and Guide (bottom). Reference image is reconstructed from k-space data without undersampling. The error is the difference between the mean, $x_\mathrm{MMSE}$, and the reference image.  The "Mean+std" is the mean highlighted with confidence interval, which indicates the reconstruction by AID is more trustworthy in the region of folding artifacts.}
  \label{fig:single}
\end{figure}

\textbf{Unfolding of aliased single-coil image:}
To investigate how the trained model, AID, reduces the folding artifacts in the reconstruction, we designed the single coil unfolding experiment. The single-channel k-space is simulated out of multichannel k-space data. The odd lines in k-space are retained, $y$. Ten samples were drawn from the posterior $p(x_1|y, x_0)$ using \cref{alg:seq} with parameters: $T=1000, \lambda=1, K=5$. The experiment was repeated using a standard diffusion model, Guide. The results are shown in \cref{fig:single}. The AID model significantly reduces the errors in the region of folding artifacts compared to the Guide model. The mean over samples, $x_\mathrm{MMSE}$, is highlighted with a confidence interval computed from the variance of samples. The highlighted mean image shows the reconstruction by AID is more trustworthy in the folding region. In general, the highlighted region lies in the folding region, where large errors remains, as we expected.

\textbf{Reconstruction from undersampled data:}
To further investigate the model's performance in reconstruction, we conducted experiments on 20 volumes from the fastMRI validation dataset where k-space data was retrospectively undersampled using various sampling masks. We created four types of sampling masks: random with autocalibration signal (ACS), random without ACS, equispaced with ACS, and equispaced without ACS. The undersampling factor is 12. Setting parameters: $T=1000, \lambda=1, K=4$ for \cref{alg:seq}, the images were reconstructed from the undersampled k-space data using the AID and Guide as prior, respectively. Another method proposed in Ref. \cite{jalal2021robust} is used as the baseline (CSGM), which uses a scored-based model (NCSNv2) from Ref. \cite{song2020improved} trained on the fastMRI dataset. 
All the reconstruction tasks are performed by sampling the posterior. The likelihood $p(y|x)$ is determined by forward model and the image prior is determined by the trained models, such as NCSNv2, Guide, and AID. This means that when the sampling method remains consistent, the performance of the reconstruction task is determined by the quality of the image prior. Our algorithm treats $p(y|x)$ in the same manner, and the key difference is the image prior.  

\begin{figure}[htbp]
  \centering
  \includegraphics[width=\textwidth]{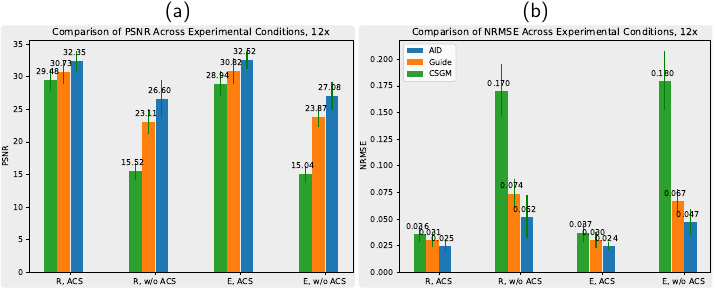}
   \caption{E: equispaced, R: random. (a): PSNR and (b): NRMSE of the images reconstructed from the twelve-times undersampled k-space data using the autoregressive diffusion model (AID), the standard diffusion model (Guide), and the baseline method CSGM. PSNR higher is better, and NRMSE lower is better.}
    \label{fig:recon1}
  \end{figure}

We used peak-signal-noise-ratio (PSNR in dB) and normalized root-mean-square error (NRMSE) to evaluate the reconstruction quality against the reference image that is reconstructed from full k-space. The comparison of metrics across  experimental conditions is illustrated in \cref{fig:recon1}. The proposed AID model outperforms the Guide and NCSNv2 in terms of PSNR and NRMSE especially in the absence of ACS, demonstrating its superior performance in image reconstruction from undersampled k-space data. The results are consistent across different undersampling factors and sampling masks, indicating the model's robustness and flexibility in handling various types of undersampled k-space data.

\begin{figure}[htbp]
  \centering
  \includegraphics[width=\textwidth]{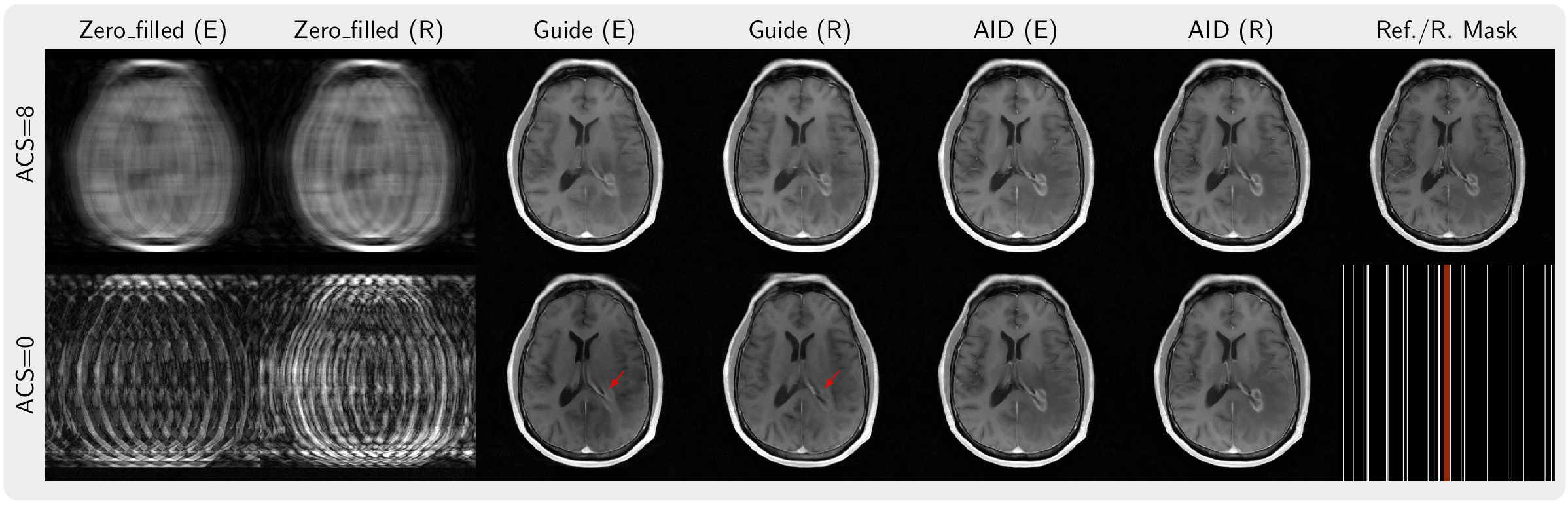}
  \caption{E: equispaced, R: random. The last column shows the reference and the random sampling mask in k-space. The red lines are autocalibration signal (ACS) and equispaced mask is not shown. Zero-filled images are computed by inverse Fourier transform of the zero-filled k-space data. The hallucinations are pointed with red arrows.}
  \label{fig:recon2}
\end{figure}

For the visual impression of the improvement by the AID model in reconstruction, we show the reconstructed images in \cref{fig:recon2} and more of them in \cref{ap:recon}. The images reconstructed using AID are more visually similar to the reference images than using Guide, even which also provides aliased-free images. Furthermore, it is worth noting that more visually notable hallucinations were introduced by the Guide model than the AID model, which means AID is more trustworthy.

\subsection{Other models for image sequence generation}
To further evaluate the model's ability to generate image sequences, we further trained two AID models on two datasets: one using brain images from autism studies called ABIDE \cite{Martino_Mol.Psychiatr._2013, Martino_Scien.Data_2017}, and the other using images from Unmanned Aerial Vehicle (UAV) view dataset \cite{delibacsouglu2022pesmod}. 
We reported the computation and model complexity in \cref{ap:com_c} for all AID models trained in this work.
We also implemented a two-stage training to improve the efficiency for training models on ABIDE and presented correspondingly generated samples in \cref{ap:two-s}.
We demonstrated the natural image sequence generation in \cref{ap:natural} and showed the sample consistency along the temporal axis in \cref{ap:consistency}.
\section{Discussion}
In this work, we propose an autoregressive image diffusion model for generating image sequences, with specific applications to accelerated MRI reconstruction. We conducted comprehensive evaluation of its performance as an image prior in reconstruction algorithms, comparing it to a standard diffusion model. Due to the learned prior information on inter-image dependencies, the proposed model outperforms the standard diffusion model across various scenarios. Our model is particularly well-suited for medical applications where image sequences are often acquired (e.g., in volumetric format) from patients in clinical practice. For instance, when different contrast images are acquired during an examination session \cite{levac2023conditional}, our model is designed to capture the relationships between these images. This enables more accurate and coherent reconstructions from undersampled k-space data using the proposed \cref{alg:seq}. Additionally, other medical imaging tasks like dynamic MRI, multi-contrast, super-resolution, and denoising could benefit from our model's ability by leveraging inter-image dependencies \cite{li2024rethinking}. Furthermore, the proposed algorithm holds great promise for facilitating the incorporation pre-existing information from other imaging modalities into MRI image reconstruction. This opens up a wide range of potential medical applications, with the potential to improve patient care and reduce healthcare costs by enabling faster and more accurate image acquisition and diagnosis.

\textbf{Privacy Issue:} As this model has the capability to generate coherent image sequences, it is crucial to consider the privacy implications associated with its use, particularly in clinical settings. The generation of such images may inadvertently expose sensitive patient information, including identifiable features such as facial characteristics. Safeguarding patient privacy must be a top priority when deploying it. We recommend that the model be used in a controlled environment where access to the generated images is restricted to authorized personnel only. Additionally, it is essential to ensure that the model is trained on anonymized data and that the generated images are not stored or shared without proper consent.

\textbf{Limitation and future work:} We did not evaluate the model on a common image dataset such as ImageNet or Cifar-10, nor did we compute metrics such as FID and Inception Score, which could be a limitation of our work.  We plan to address these limitations in future work by running the model on a large dataset and comparing it with other state-of-the-art models. Additionally, given the model's suitability for modeling image sequences, it is worth exploring its potential for optimizing MRI k-space acquisition strategies, as the acquisition process constitutes a sequence of operations.

\section{Conclusion}
The proposed autoregressive image diffusion model offers an approach to generating image sequences, with significant potential as a trustworthy prior in accelerated MRI reconstruction. In various experiments, it outperforms the standard diffusion model in terms of both image quality and robustness by taking the advantage of the prior information on inter-image dependencies. 

\section*{Acknowledgements}
This work was supported by DZHK (German Centre for Cardiovascular Research) funding code: 81Z0300115. We acknowledge funding by the "Niedersächsisches Vorab" funding line of the Volkswagen Foundation. This work was supported by the Federal Ministry of Education and Research (BMBF), Germany under the AI service center KISSKI (grant no. 01IS22093A).

\bibliographystyle{unsrtnat}
\bibliography{ref}

\appendix

\newpage
\section{Loss function derivation}
\label{ap:loss}
Below is a derivation of \cref{eq:dn}, the reduced variance variational bound for diffusion models. This adapted from \citet{sohl2015deep} and \citet{ho2020denoising}. We include it here only for completeness. In the forward process, $x_n^{t}$ and $x_{<n}^0$ are conditionally independent given $x_n^{t-1}$.
\begin{align}
  L &= \mathbb{E}_{q} \left[ -\log p ( {x}^{T}_n| x_{<n}^0) - \sum_{t > 1} \log \frac{p_{\theta} ( {x}^{t-1}_n | {x}^{t}_n, x_{<n}^0 )}{q ( {x}^{t}_n | {x}^{t-1}_n, x_{<n}^0)} - \log \frac{p_{\theta} ( {x}^{0}_n | {x}^{1}_n, x_{<n}^0)}{q ( {x}^{1}_n | {x}^{0}_n,x_{<n}^0)} \right] \\
  &= \mathbb{E}_{q} \left[ -\log p ( {x}^{T}_n|x_{<n}^0) - \sum_{t > 1} \log{\frac{p_\theta ( {x}^{t-1}_n | {x}^{t}_n, x_{<n}^0)}{q ( {x}^{t-1}_n | {x}^{t}_n,x^0_n,x_{<n}^0 )}} \cdot{\frac{q ( {x}^{t-1}_n | {x}_n^{0} )}{q ( {x}^{t}_n| {x}^{0}_n )}} - \log{\frac{p_{\theta} ( {x}^{0}_n | {x}^{1}_n, x_{<n}^0)}{q ( {x}^{1}_n| {x}^{0}_n, {x}^{0}_{<n} )}} \right] \\
   &= \mathbb{E}_{q} \left[ -\log{\frac{p ( {x}^{T}_n|x_{<n}^0)}{q ( {x}^{T}_n | {x}^{0}_n,x_{<n}^0 )}} - \sum_{t > 1} \log{\frac{p_{\theta} ( {x}^{t-1}_n | {x}^{t}_n,x_{<n}^0)}{q ( {x}^{t-1}_n | {x}^{t}_n, {x}^{0}_n,x_{<n}^0 )}} - \log \frac{p_{\theta} ( {x}^{0}_n | {x}^{1}_n,x_{<n}^0 )}{q(x^1_n|x^0_n,x_{<n}^0)} \right] \\
   &= \mathbb{E}_{q} \Bigg[ D_{\mathrm{KL}} ( q ( {x}^{T}_n | {x}^{0}_n,x_{<n}^0 ) \parallel p ( {x}^{T}_n|x_{<n}^0 ) ) + \sum_{t > 1} D_{\mathrm{KL}} ( q ( {x}^{t-1}_n | {x}^{t}_n, {x}^{0}_n ) \parallel p_{\theta} ( {x}^{t-1}_n | {x}^{t}_n,x_{<n}^0 ) )\\
   & \phantom{text}- \log p_{\theta} ( {x}^{0}_n | {x}^{1}_n, x_{<n}^0 ) \Bigg]
\end{align}

\section{Posterior derivation}
\label{ap:post}
When samples drawn from the posterior started from the standard Gaussian noise, with \cref{eq:pos1} we have
\begin{align}
  p(x_n^t|y_n, x_{<n}^0) &\propto p(y_n|x_n^t)p(x_n^t|x_{<n}^0)~
 \label{eq:pos2}
\end{align}
for all the reverse time steps.
Because 
\begin{align}
  p({x}^{t}_n|x_{<n}^0)=\int p({x}_n^{t}|{ x}_n^{t+1},x_{<n}^0)p({ x}_n^{t+1})d{x}_n^{t+1}
\end{align}
and 
\begin{align}
  \int p(x_n^t|x_n^{t+1}, y_n, x_{<n}^0)p(x_n^{t+1})dx_n^{t+1}& = p(x_n^t|y_n,x_{<n}^0)~,\\
  &= \frac{p(y_n|x_n^t)p(x_n^t|x_{<n}^0)}{p(y_n)}~,
\end{align}
then we have
\begin{align}
  \int p(x_n^t|x_n^{t+1}, y_n, x_{<n}^0)p(x_n^{t+1})dx_n^{t+1} &= \frac{p(y_n|x_n^t)}{p(y_n)}\cdot\int p({x}_n^{t}|{ x}_n^{t+1},x_{<n}^0)p({ x}_n^{t+1})d{x}_n^{t+1}.
\end{align}
Therefore, we have
\begin{align}
  p(x_n^{t}|x_n^{t+1}, y_n, x_{<n}^0) = \frac{p(y_n|x_n^t)p(x_n^{t}|x_n^{t+1}, x_{<n}^0)}{p(y_n)}~.
\end{align}
$p(y_n)$ is a constant for evidence. Then with gradient based method, the posterior $p(x_n^{t}|x_n^{t+1}, y_n, x_{<n}^0)$ is sampled from the likelihood $p(y_n|x_n^t)$ and the reverse process $p(x_n^{t}|x_n^{t+1}, x_{<n}^0)$, 

\section{Likelihood function for k-space}
The autocalibration signal (ACS) region are lines through the center of k-space, however, are fully sampled. The sensitivity of a coil is a spatial profile
that describes the receiving field that induces signals in the coil. The simultaneous data acquisition, with each coil's sensitivity corresponding to a different subregion, leads to a complete
image without aliasing artifacts.
\label{ap:lik}
\begin{figure}[htbp]
  \centering
  \includegraphics[width=\textwidth]{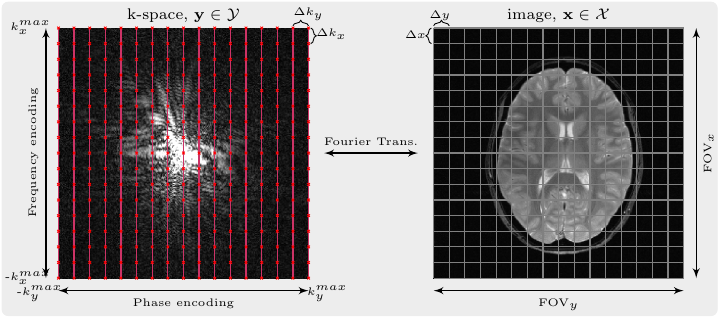}
  \caption{The relationship between k-space and image. The Nyquist theorem states that the sampling rate must be at least twice the highest frequency component in the signal.}
  \label{fig:nyq}
\end{figure}

\begin{figure}[htbp]
  \centering
  \includegraphics[width=\textwidth]{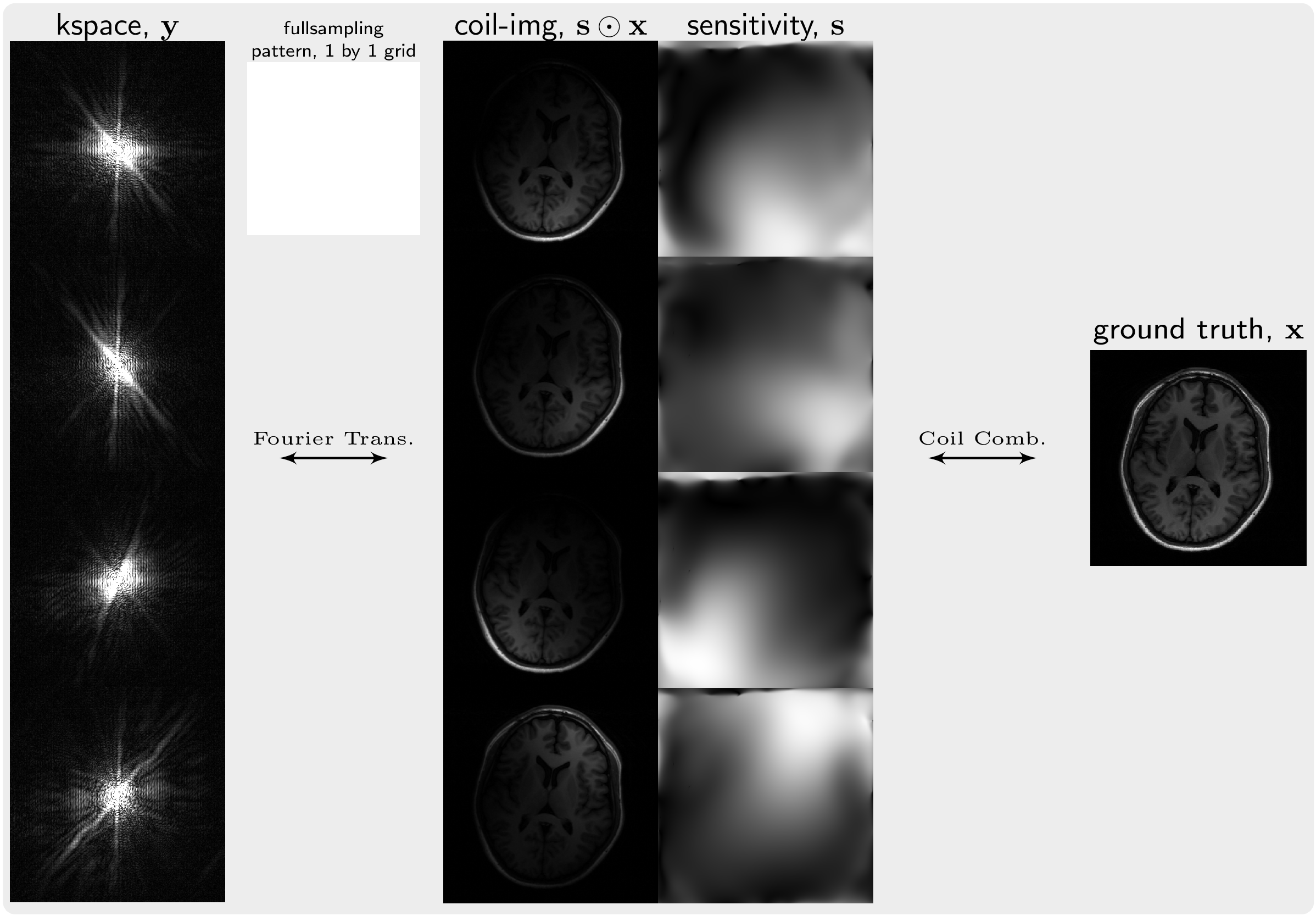}
  \caption{The signal detected by a coil is weighted by its local coil profile, which is called sensitivities and imposes weights on the
  signal intensity. Consequently, it causes dark and bright regions in coil images. The ground truth image is the combination of all coil images.}
\end{figure}

\section{Cardiac samples}
\label{ap:cardiac}

\begin{figure}[H]
  \centering
  \includegraphics[width=\textwidth]{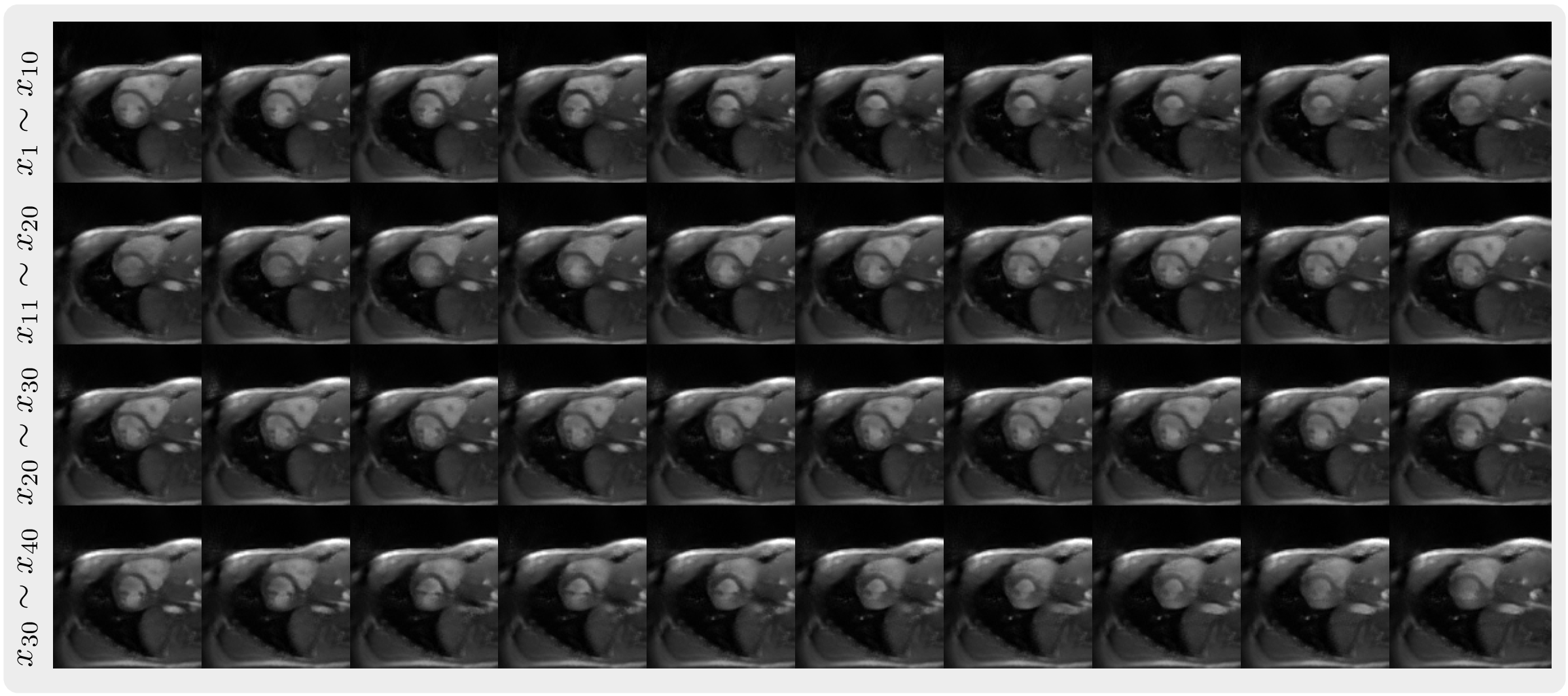}
  \caption{Prospective samples from the model trained on cardiac dataset.}
\end{figure}
\begin{figure}[H]
  \centering
  \includegraphics[width=\textwidth]{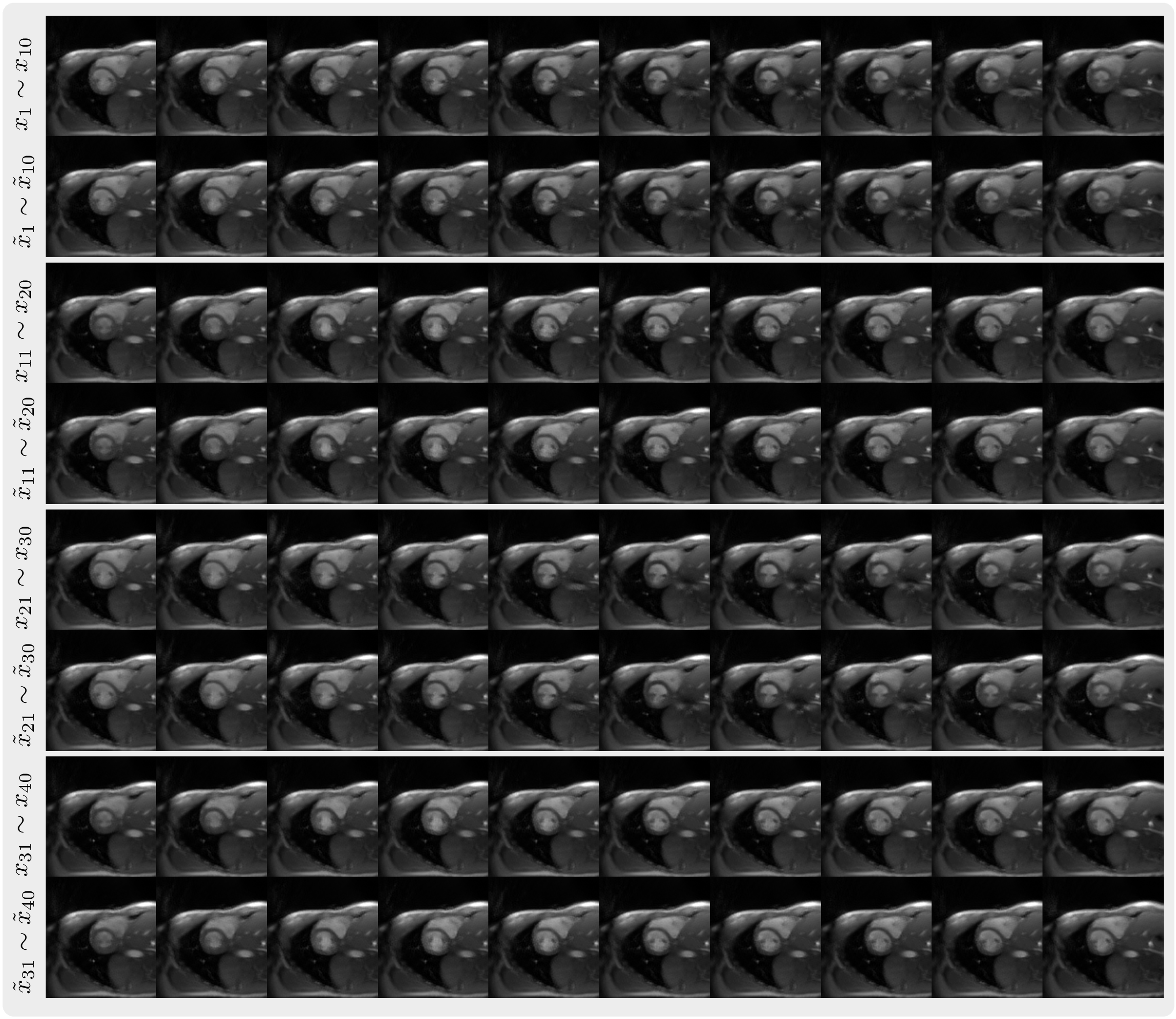}
  \caption{Retrospective samples from the model trained on cardiac dataset.}
\end{figure}

\section{Reconstruction from undersampled data}
\label{ap:recon}
\begin{figure}[H]
  \centering
  \includegraphics[width=\textwidth]{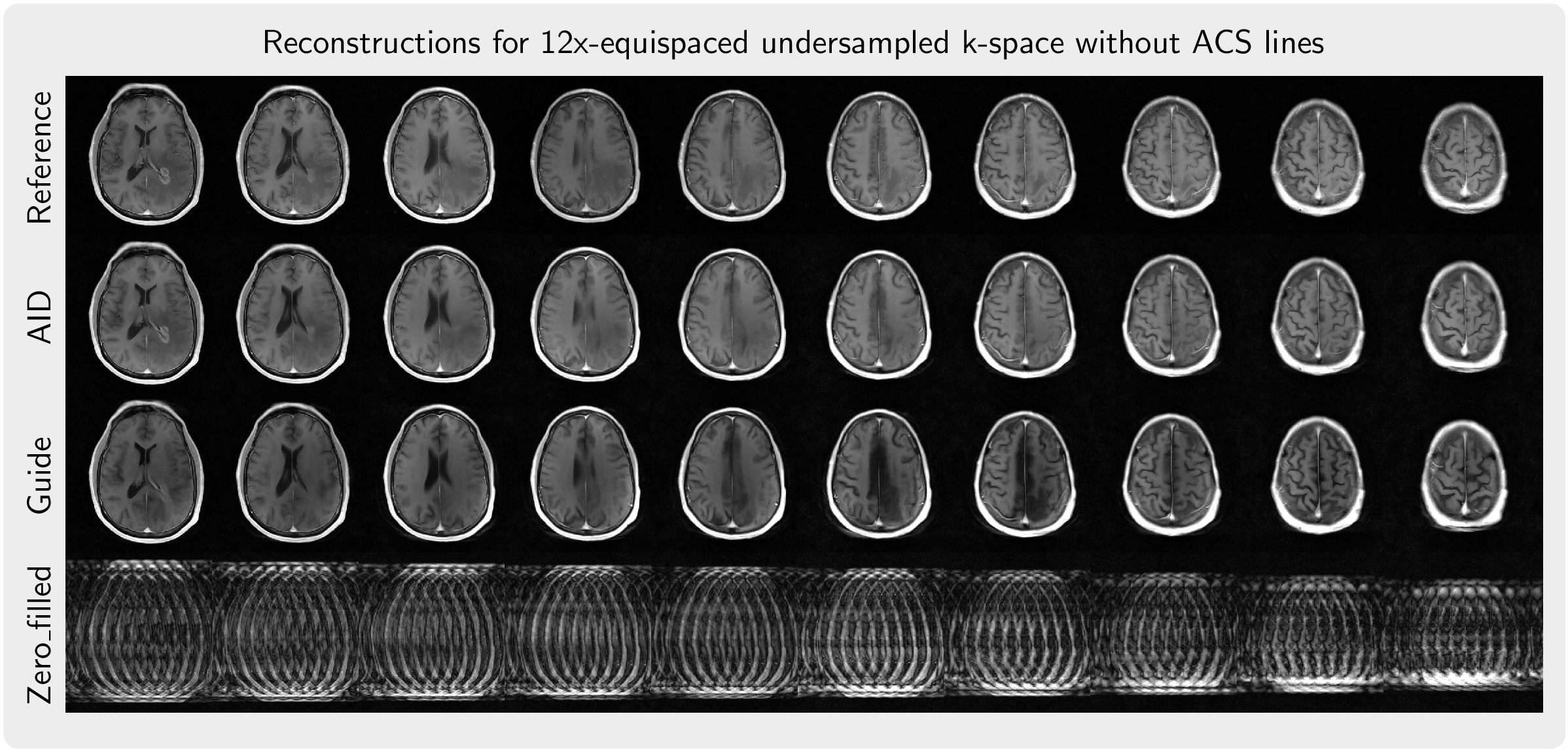}
\end{figure}
\begin{figure}[H]
  \centering
  \includegraphics[width=\textwidth]{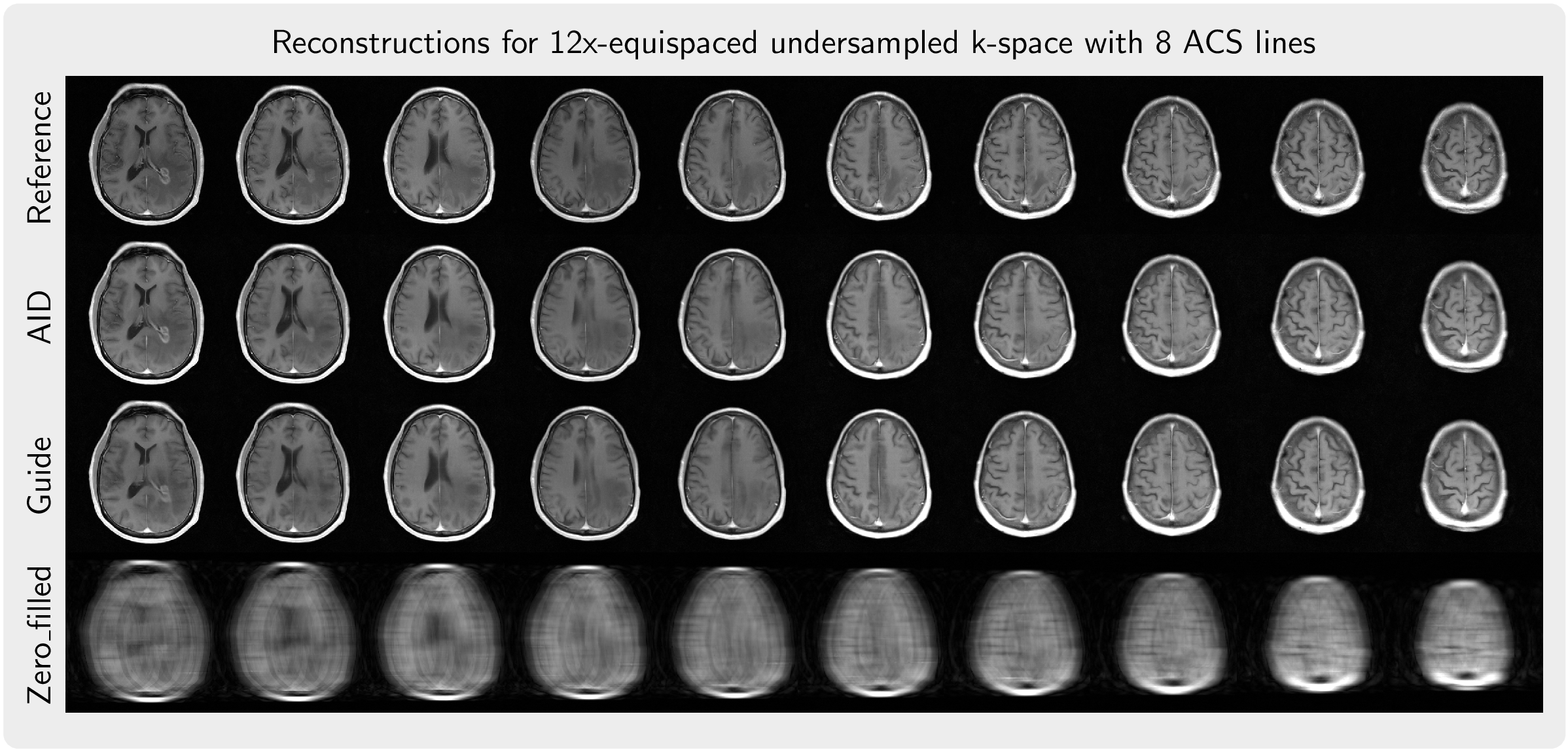}
\end{figure}

\begin{figure}[H]
  \centering
  \includegraphics[width=\textwidth]{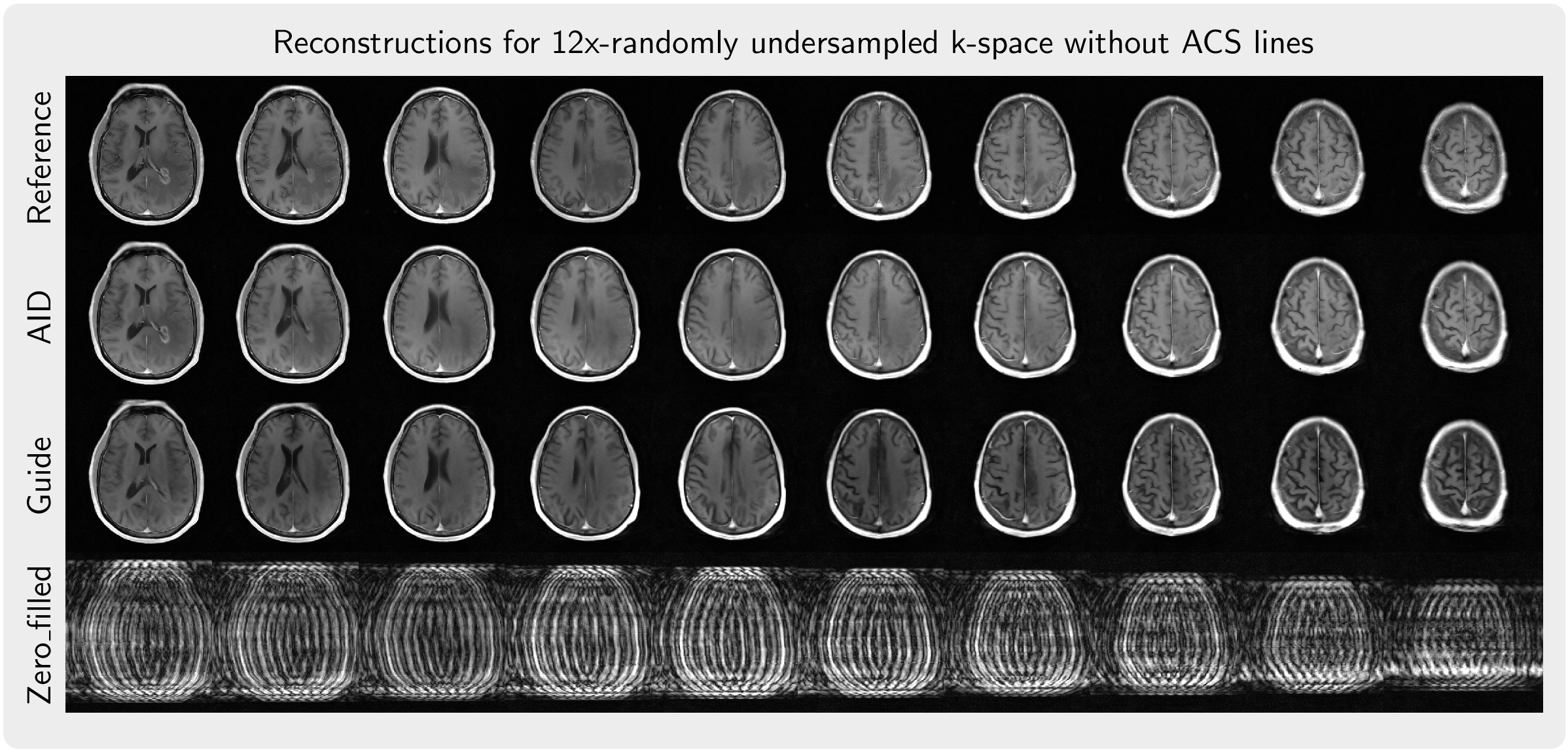}
\end{figure}

\begin{figure}[H]
  \centering
  \includegraphics[width=\textwidth]{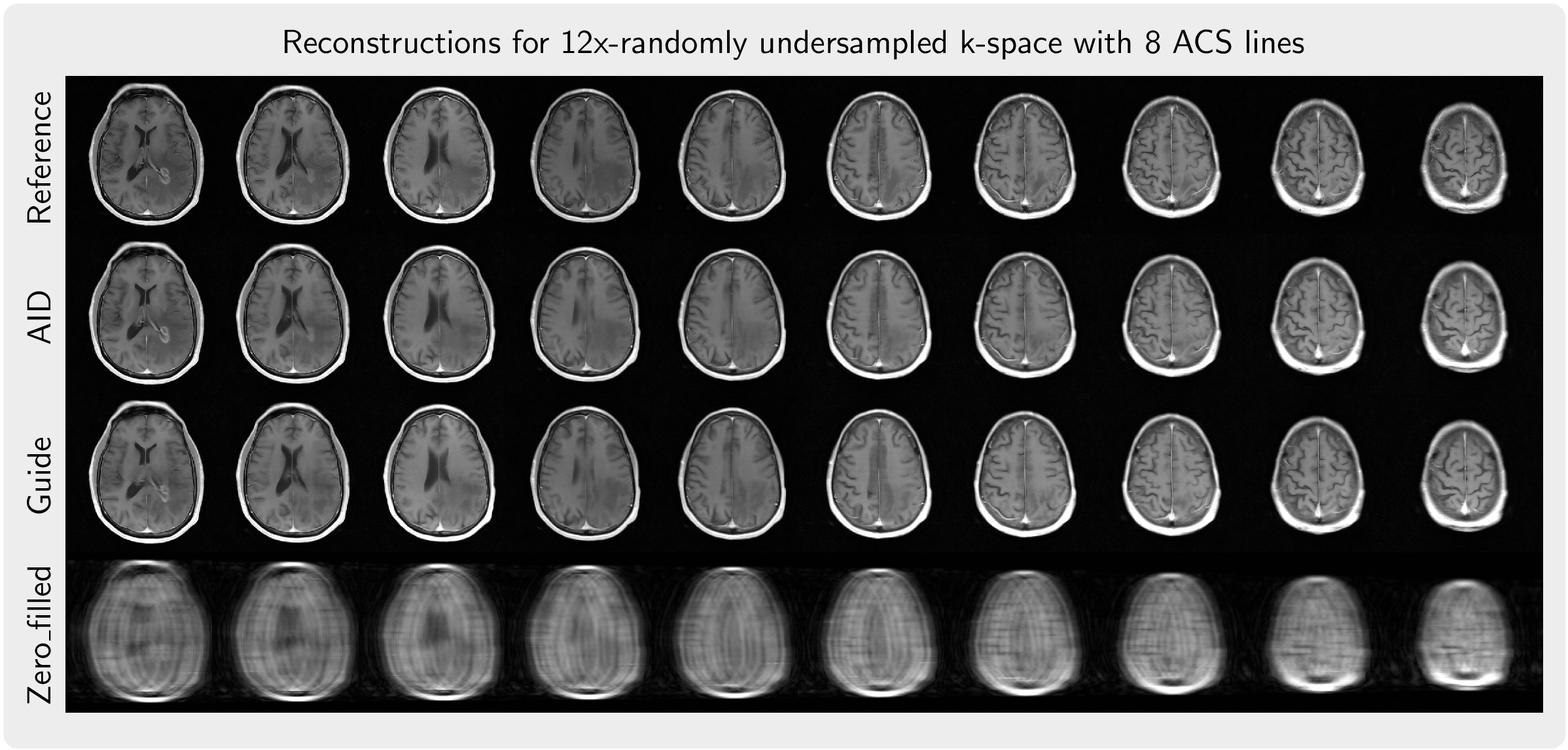}
\end{figure}

\section{Computation and model complexity}
\label{ap:com_c}
Table 1  presents the computation needed to train the AID models on four different datasets with different model complexities.
\begin{itemize}[left=0em]
\tightlist[1pt]
\item Dataset: the name of the dataset (fastMRI, cardiac, ABIDE, UAV).
\item Length: the length of the image sequence.
\item Image size: the dimensions of the images in the dataset.
\item Latent: the latent space representation used for the model, with options like VQVAE, Autoencoder-KL, or None.
\item Two-stage: a boolean indicating whether a two-stage training process was used. Two-stage training is explained in the following section.
\item Parameters: the number of parameters in the model.
\item Train steps/s: training speed in steps per second.
\item Inference (it/s): inference speed in iterations per second.
\end{itemize}

\begin{table}[htbp]
  \caption{Datasets and computational resources used to train the four different AID models.\label{tab:priors}}
  \vspace{-5pt}
  \def\arraystretch{1.1}
  \resizebox{\textwidth}{!}{
  \centering
  \begin{tabular}{c@{\hskip 10pt}c @{\hskip 10pt}c @{\hskip 10pt}c@{\hskip 10pt}c @{\hskip 10pt}c @{\hskip 10pt}c@{\hskip 10pt}c}%
    \toprule
      Dataset & Length & Image size & Latent  & Two-stage & Parameters         & Train steps/s & Inference (it/s) \\ \hline
      fastMRI & 10     & 320 &  None          & False     & \textasciitilde139M & \textasciitilde 1.31& \textasciitilde 10.07\\
      cardiac & 42     & 256 & VQVAE, 4x      & False     & \textasciitilde26M & \textasciitilde 1.27 & \textasciitilde 20.20\\
      ABIDE   & 46     & 128 & None           & True      & \textasciitilde36M & \textasciitilde 0.89 & \textasciitilde 4.10\\
      UAV     & 70     & 256 & Autoencoder-KL, 8x & True  & \textasciitilde83M & \textasciitilde 1.05 & \textasciitilde 39.60\\
\bottomrule 
\end{tabular}}
\end{table}

\section{Two-stage training}
\label{ap:two-s}
\begin{figure}[htbp]
  \centering
  \includegraphics[width=\textwidth]{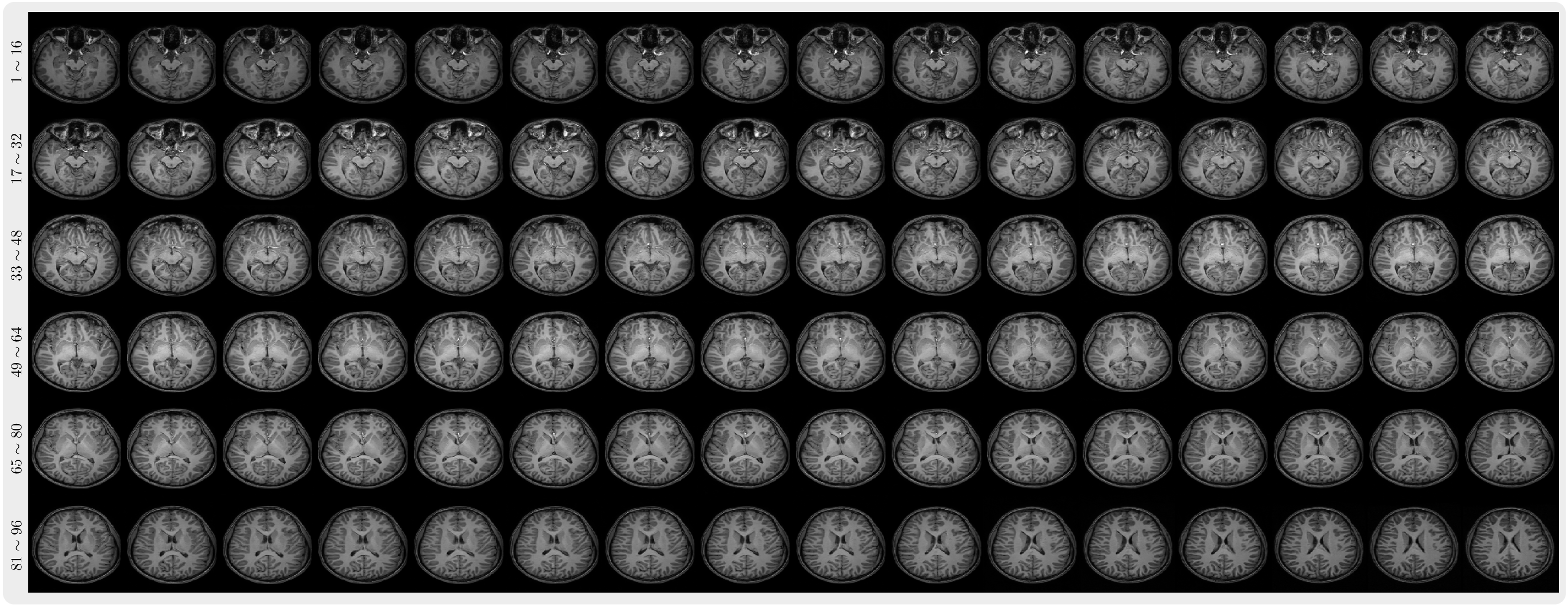}
  \vspace{-1.5em}
  \caption{The generated ABIDE image sequence captures the changes in the brain structure.}
  \label{fig:abide}
\end{figure}

We implemented a two-stage training process to improve training efficiency. In the first stage, we trained the U-net model. In the second stage, we trained the temporal-spatial conditioning block with the pre-trained U-net model frozen. By doing so, we are able to train an AID model on ABIDE dataset, where the image sequence has a dimension of 46\texttimes128\texttimes128 after preprocessing. The generated image squeence is shown in \cref{fig:abide}.

\section{Natural image sequence generation}
\label{ap:natural}
\begin{figure}[htbp]
  \centering
  \includegraphics[width=\textwidth]{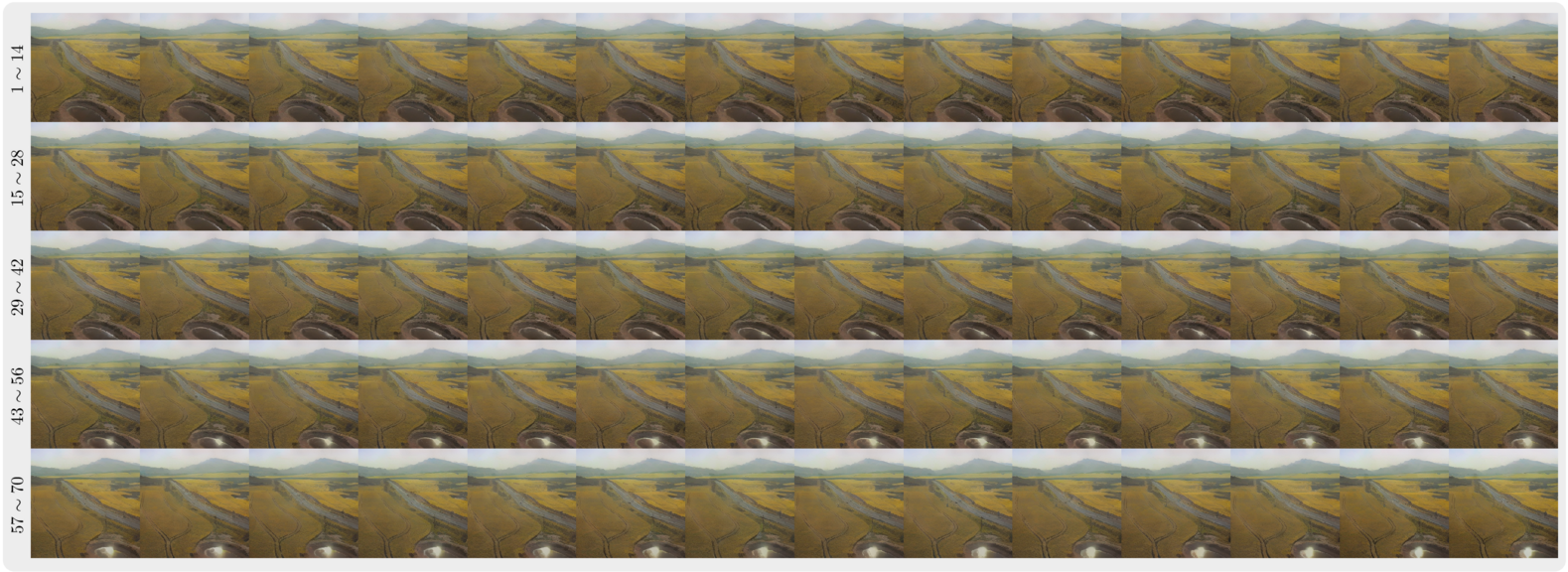}
  \vspace{-1.5em}
  \caption{The light changes in the water surface are captured in the generated UAV image sequence.}
  \label{fig:uav}
\end{figure}

We trained an AID model on an UAV dataset in the latent space and generated images using the trained model. The generated images are displayed in \cref{fig:uav}. The generated images demonstrate the effectiveness of the proposed method in learning sequentially coherent natural images generation. Each frame in Figure 2 shows an aerial view of a rural landscape with roads and/or a water pond.

\section{Sample consistency along the temporal axis}
\label{ap:consistency}
\begin{figure}[htbp]
    \centering
    \includegraphics[width=\textwidth]{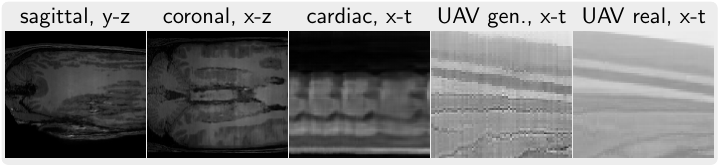}
    \vspace{-1em}
    \caption{Temporal consistency of the images generated by AID models trained on different datasets. The first two columns show the sagittal and coronal view of brain image sequence. The x-t plane of cardiac image and UAV sequence are shown in the last three columns.}
    \label{fig:consistency}
\end{figure}

\cref{fig:consistency} shows the sample consistency along the temporal (or z) axis. Columns 1 and 2: Show sagittal and coronal views of a brain image sequence. These images appear to be medical scans with clearly stretched anatomical structures.
Column 3: Displays the x-t plane of a cardiac image sequence. This displays the heart's activity over time and shows the diastolic and systolic phases from left to right.
Columns 4 and 5: Show the x-t plane of a UAV image sequence, both generated and real. These images show the change in aerial views of a landscape over time. The generated x-t plane are generally consistent with the real x-t plane images but suffer from the striped artifacts.

\section{VQVAE configuration for cardiac dataset}
\label{ap:vqvae}
The VQVAE is trained on the cardiac dataset to generate the latent space for the training of the autoregressive diffusion model, using the official implementation\footnote{https://github.com/CompVis/taming-transformers.git}. The VQVAE is trained with the following configuration:
\begin{verbatim}
base_learning_rate: 4.5e-06
params:
  embed_dim: 3
  n_embed: 8192
  ddconfig:
    double_z: false
    z_channels: 3
    resolution: 256
    in_channels: 3
    out_ch: 3
    ch: 128
    ch_mult: [1, 2, 4]
    num_res_blocks: 2
    attn_resolutions: []
    dropout: 0.0
  lossconfig:
    target: losses.vqperceptual.VQLPIPSWithDiscriminator
    params:
      disc_conditional: false
      disc_in_channels: 3
      disc_start: 30001
      disc_weight: 0.8
      codebook_weight: 1.0
\end{verbatim}

\section{3D volume generation}
\label{ap:3d}
To further improve the model's ability to generate 3D volumes, the position embedding is added to the third dimension - z of the volume. This allows the model trained on the  ABIDE dataset to have better consistency along the z-axis.

\begin{figure}[H]
  \centering
  \includegraphics[width=0.45\textwidth, angle=90]{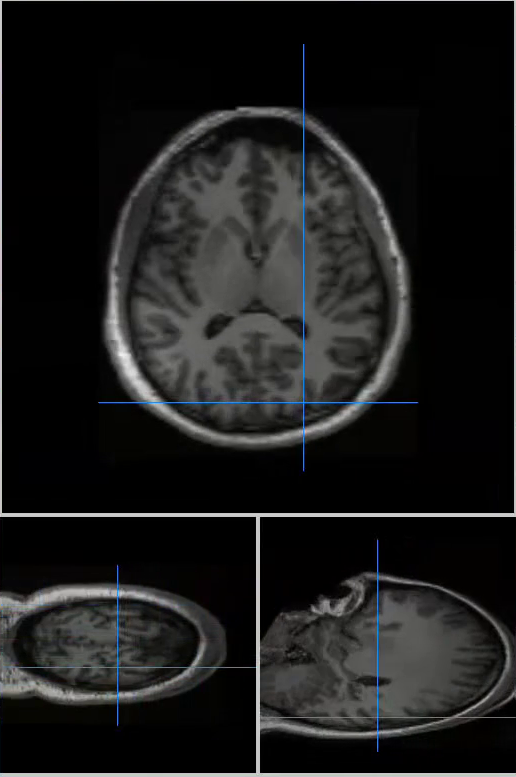}
  \caption{The generated 3D volumes from the model trained on the ABIDE dataset. The diffusion process is applied on the transverse plane (x-y) (c.f. the image on the left) and the autoregressive process is applied on the z-axis.}
  \label{fig:3d}
\end{figure}

\ifthenelse{\boolean{arxiv}}{}{
\newpage
\section*{NeurIPS Paper Checklist}

%%% BEGIN INSTRUCTIONS %%%
The checklist is designed to encourage best practices for responsible machine learning research, addressing issues of reproducibility, transparency, research ethics, and societal impact. Do not remove the checklist: {\bf The papers not including the checklist will be desk rejected.} The checklist should follow the references and follow the (optional) supplemental material.  The checklist does NOT count towards the page
limit. 

Please read the checklist guidelines carefully for information on how to answer these questions. For each question in the checklist:
\begin{itemize}
    \item You should answer \answerYes{}, \answerNo{}, or \answerNA{}.
    \item \answerNA{} means either that the question is Not Applicable for that particular paper or the relevant information is Not Available.
    \item Please provide a short (1–2 sentence) justification right after your answer (even for NA). 
   % \item {\bf The papers not including the checklist will be desk rejected.}
\end{itemize}

{\bf The checklist answers are an integral part of your paper submission.} They are visible to the reviewers, area chairs, senior area chairs, and ethics reviewers. You will be asked to also include it (after eventual revisions) with the final version of your paper, and its final version will be published with the paper.

The reviewers of your paper will be asked to use the checklist as one of the factors in their evaluation. While "\answerYes{}" is generally preferable to "\answerNo{}", it is perfectly acceptable to answer "\answerNo{}" provided a proper justification is given (e.g., "error bars are not reported because it would be too computationally expensive" or "we were unable to find the license for the dataset we used"). In general, answering "\answerNo{}" or "\answerNA{}" is not grounds for rejection. While the questions are phrased in a binary way, we acknowledge that the true answer is often more nuanced, so please just use your best judgment and write a justification to elaborate. All supporting evidence can appear either in the main paper or the supplemental material, provided in appendix. If you answer \answerYes{} to a question, in the justification please point to the section(s) where related material for the question can be found.

IMPORTANT, please:
\begin{itemize}
    \item {\bf Delete this instruction block, but keep the section heading ``NeurIPS paper checklist"},
    \item  {\bf Keep the checklist subsection headings, questions/answers and guidelines below.}
    \item {\bf Do not modify the questions and only use the provided macros for your answers}.
\end{itemize}

%%% END INSTRUCTIONS %%%

\begin{enumerate}

\item {\bf Claims}
    \item[] Question: Do the main claims made in the abstract and introduction accurately reflect the paper's contributions and scope?
    \item[] Answer: \answerYes{} % Replace by \answerYes{}, \answerNo{}, or \answerNA{}.
    \item[] Justification: See the last paragraph of the introduction in Page 2. %\justificationTODO{}

\item {\bf Limitations}
    \item[] Question: Does the paper discuss the limitations of the work performed by the authors?
    \item[] Answer: \answerYes{} % Replace by \answerYes{}, \answerNo{}, or \answerNA{}.
    \item[] Justification: See the last paragraph of discussion in Page 10. %\justificationTODO{}

\item {\bf Theory Assumptions and Proofs}
    \item[] Question: For each theoretical result, does the paper provide the full set of assumptions and a complete (and correct) proof?
    \item[] Answer: \answerYes{} % Replace by \answerYes{}, \answerNo{}, or \answerNA{}.
    \item[] Justification: See Appendix A and B %\justificationTODO{}

    \item {\bf Experimental Result Reproducibility}
    \item[] Question: Does the paper fully disclose all the information needed to reproduce the main experimental results of the paper to the extent that it affects the main claims and/or conclusions of the paper (regardless of whether the code and data are provided or not)?
    \item[] Answer: \answerYes{} % Replace by \answerYes{}, \answerNo{}, or \answerNA{}.
    \item[] Justification: The paper provides all the information needed to reproduce the main experimental results, include data, training details, algorithm details, and evaluation metrics in the section of experiments. %\justificationTODO{}

\item {\bf Open access to data and code}
    \item[] Question: Does the paper provide open access to the data and code, with sufficient instructions to faithfully reproduce the main experimental results, as described in supplemental material?
    \item[] Answer: \answerYes{} % Replace by \answerYes{}, \answerNo{}, or \answerNA{}.
    \item[] Justification: The code is available at \url{https://github.com/mrirecon/aid}.

\item {\bf Experimental Setting/Details}
    \item[] Question: Does the paper specify all the training and test details (e.g., data splits, hyperparameters, how they were chosen, type of optimizer, etc.) necessary to understand the results?
    \item[] Answer: \answerYes{} % Replace by \answerYes{}, \answerNo{}, or \answerNA{}.
    \item[] Justification: The paper provides all the necessary details to understand the results in the section of experiments. %\justificationTODO{}

\item {\bf Experiment Statistical Significance}
    \item[] Question: Does the paper report error bars suitably and correctly defined or other appropriate information about the statistical significance of the experiments?
    \item[] Answer: \answerYes{} % Replace by \answerYes{}, \answerNo{}, or \answerNA{}.
    \item[] Justification: The paper reports error bar in Figure 5. %\justificationTODO{}

\item {\bf Experiments Compute Resources}
    \item[] Question: For each experiment, does the paper provide sufficient information on the computer resources (type of compute workers, memory, time of execution) needed to reproduce the experiments?
    \item[] Answer: \answerYes{} % Replace by \answerYes{}, \answerNo{}, or \answerNA{}.
    \item[] Justification: See Section 3.1
    
\item {\bf Code Of Ethics}
    \item[] Question: Does the research conducted in the paper conform, in every respect, with the NeurIPS Code of Ethics \url{https://neurips.cc/public/EthicsGuidelines}?
    \item[] Answer: \answerYes{} % Replace by \answerYes{}, \answerNo{}, or \answerNA{}.
    \item[] Justification: See \textbf{Privacy Issue} in the section of discussion. %\justificationTODO{}

\item {\bf Broader Impacts}
    \item[] Question: Does the paper discuss both potential positive societal impacts and negative societal impacts of the work performed?
    \item[] Answer: \answerYes{} % Replace by \answerYes{}, \answerNo{}, or \answerNA{}.
    \item[] Justification: See \textbf{Privacy Issue} in the section of discussion.
    
\item {\bf Safeguards}
    \item[] Question: Does the paper describe safeguards that have been put in place for responsible release of data or models that have a high risk for misuse (e.g., pretrained language models, image generators, or scraped datasets)?
    \item[] Answer: \answerNA{} % Replace by \answerYes{}, \answerNo{}, or \answerNA{}.
    \item[] Justification: The paper does not release data or models that have a high risk for misuse. %\justificationTODO{}

\item {\bf Licenses for existing assets}
    \item[] Question: Are the creators or original owners of assets (e.g., code, data, models), used in the paper, properly credited and are the license and terms of use explicitly mentioned and properly respected?
    \item[] Answer: \answerYes{} % Replace by \answerYes{}, \answerNo{}, or \answerNA{}.
    \item[] Justification: The creators or original owners of assets are properly credited and the license and terms of use are explicitly mentioned and properly respected.

\item {\bf New Assets}
    \item[] Question: Are new assets introduced in the paper well documented and is the documentation provided alongside the assets?
    \item[] Answer: \answerNA{} % Replace by \answerYes{}, \answerNo{}, or \answerNA{}.
    \item[] Justification: The paper does not submit new assets. %\justificationTODO{}

\item {\bf Crowdsourcing and Research with Human Subjects}
    \item[] Question: For crowdsourcing experiments and research with human subjects, does the paper include the full text of instructions given to participants and screenshots, if applicable, as well as details about compensation (if any)? 
    \item[] Answer: \answerNo{} % Replace by \answerYes{}, \answerNo{}, or \answerNA{}.
    \item[] Justification: This work does not involve crowdsourcing nor research with human subjects. %\justificationTODO{}

\item {\bf Institutional Review Board (IRB) Approvals or Equivalent for Research with Human Subjects}
    \item[] Question: Does the paper describe potential risks incurred by study participants, whether such risks were disclosed to the subjects, and whether Institutional Review Board (IRB) approvals (or an equivalent approval/review based on the requirements of your country or institution) were obtained?
    \item[] Answer: \answerYes{} % Replace by \answerYes{}, \answerNo{}, or \answerNA{}.
    \item[] Justification: This paper paper use publicly released fastMRI dataset and have been approved by the local IRB. %\justificationTODO{}

\end{enumerate}}

\end{document}